\documentclass{article}
\usepackage[nonatbib]{arxiv}
\usepackage[utf8]{inputenc}
\usepackage{lmodern}
\usepackage[T1]{fontenc}
\usepackage[numbers,comma,square,sort&compress]{natbib}
\usepackage{hyperref}
\hypersetup{pdfborder={0 0 0},
            colorlinks=true,
            urlcolor=blue,
            breaklinks=true,
            extension = }
\newcommand{\dantondef}[2]{
        \href{https://niess.github.io/danton-docs/\#HEAD/#1}{\texttt{#2}}}
\newcommand{\dantonfun}[2]{
        \href{https://niess.github.io/danton-docs/\#HEAD/#1/#2}{\texttt{#2}}}
\usepackage[document]{ragged2e}
\usepackage{graphicx}
\usepackage{float}
\usepackage[caption=false]{subfig}
\usepackage{enumerate}
\usepackage{booktabs}
\usepackage{textcomp}

\title{DANTON: a Monte-Carlo sampler of $\tau$ from $\nu_\tau$
interacting with the Earth}

\author{
  Valentin~Niess \\
  Université Clermont Auvergne, CNRS/IN2P3, LPC \\
  Clermont-Ferrand, France \\
  \texttt{niess@in2p3.fr} \\
  \And Olivier Martineau-Huynh \\
  LPNHE, Université Pierre et Marie Curie, Université Paris Diderot, CNRS/IN2P3 \\
  Paris, France \\
  \texttt{martineau@lpnhe.in2p3.fr} \\
}

\begin{document}
\maketitle

\begin{abstract}
A preliminary version (\texttt{v0.2.1}) of the DANTON Monte-Carlo package is
presented. DANTON allows the exclusive sampling of (decaying) $\tau$ generated
by $\nu_\tau$ interactions with the Earth. The particles interactions with
matter are simulated in detail, including transverse scattering. Detailed
topography data of the Earth can be used as well. Yet, high Monte-Carlo
efficiency is achieved by using a Backward Monte-Carlo technique. Some
validation results are provided.
\end{abstract}

\section{Introduction}
\justify
The present document is a brief technical note describing the DANTON (DecAyiNg
Taus frOm Neutrinos) Monte-Carlo sampler. DANTON is used for the simulation of
the Giant RAdio Neutrino Detector (GRAND, see e.g.~\citet{GRAND:ICRC2017}). It
is a generator of ultra relativistic $\tau$ originating from the interactions of
$\nu_\tau$ of cosmic origin with the Earth. Motivations for this work, as well
as a comprehensive description of the topic, can be found
in~\citet{Schoorlemmer:2018}. Comparisons to the
NuTauSim\,\cite{NuTauSim:GitHub} Monte-Carlo code are also provided hereafter.

The main specificity of DANTON are the following:
\begin{enumerate}[(i)]
        \item DANTON can be operated in both forward (classical) or Backward
        Monte-Carlo (BMC) mode. The latter is done following the approach
        described in~\citet{Niess:2018}. The BMC mode allows DANTON to
        exclusively sample a given final $\tau$ state.

        \item DANTON implements a detailed simulation of Physical processes,
        i.e. energy losses are stochastic. The transport of $\nu_\tau$ is done
        with ENT\,\cite{ENT:GitHub}. The cross-sections for $\nu$ Deep Inelastic
        Scattering (DIS) are computed by ENT from tabulated Parton Distribution
        Functions (PDF). The transport of $\tau$ is done
        with~PUMAS\,\cite{PUMAS:pages}. The $\tau$ energy loss is read from
        material tables. Decays of $\tau$ are simulated with
        ALOUETTE\,\cite{ALOUETTE:GitHub}, a C wrapper to TAUOLA\,\cite{TAUOLA}
        providing BMC decays.

        \item DANTON can integrate a detailed topography from global models,
        e.g. ASTER\,\cite{ASTER} or SRTMGL1\,\cite{SRTMGL1}. This is done with
        the TURTLE library\,\cite{TURTLE:GitHub}. It also includes a US standard
        atmospheric model.
\end{enumerate}

The DANTON package has been implemented in C99. Its source code is available
from GitHub\,\cite{DANTON:GitHub}. Its functionalities are exposed to the
end user as a dedicated library, \texttt{libdanton}, following an Object
Oriented (OO) approach. A documentation of the library Application Programming
Interface (API) is available online\,\cite{DANTON:API}. The API allows to
efficiently integrate DANTON in higher level simulation schemes. An executable
is provided as well: \texttt{danton}, encapsulating \texttt{libdanton}. It
allows to configure and run DANTON simulations according to steering files in
JSON format.

\section{The DANTON library}

\subsection{Initialisation and finalisation}
Prior to any usage, \texttt{libdanton} must be initialised. This is done by
calling the \dantonfun{group/danton}{danton\_initialise} function. The
initialisation allows to provide data sets for the Physics. The PDF can be
selected. It must be in Les Houches Accords Grid~1 format (\texttt{lhagrid1}).
The $\tau$ energy loss tables for PUMAS can be selected as well. Their must
conform to the PDG\,\cite{PhysRevD.98.030001} format. If a \texttt{NULL}
argument is provided, then a default data set is loaded. This is
CT14\_NLO\,\cite{CT14} for PDFs or self generated tables for PUMAS, with
Photonuclear interactions modelled according to \citet{Dutta:2001}. For
multithreaded usage, one must also provide a couple of
\dantonfun{group/callback}{danton\_lock\_cb} callbacks in order to \emph{lock}
and \emph{unlock} access to critical data sections. These callbacks might for
example manage a mutex using pthread.

Finalising the library is done with the
\dantonfun{group/danton}{danton\_finalise} function. Any memory allocated by a
\texttt{libdanton} function is automatically released at finalisation. Whenever
one wants to explicitly release a \texttt{libdanton} object, one must call the
generic \dantonfun{group/danton}{danton\_destroy} function instead of the
standard library function: \texttt{free}. This will correctly notify
\texttt{libdanton} that the object has been released. Note that some objects
require a specific destructor to be called, e.g. the \texttt{danton\_context}.

\subsection{Error handling}
Whenever an error occurs when calling a \texttt{libdanton} function, it will
return \texttt{EXIT\_FAILURE} or a \texttt{NULL} pointer. A short description of
the error is pilled up in an internal stack. It can be retrieved using the
\dantonfun{group/error}{danton\_error\_pop} function. The numbers of errors in
the stack is provided by the \dantonfun{group/error}{danton\_error\_count}
function. It is also possible to push an error to the stack, using the
\dantonfun{group/error}{danton\_error\_push} function.

\subsection{Setting the geometry}
Before running any simulation, the geometry must be defined. The Earth model is
specified with the \dantonfun{group/earth}{danton\_earth\_model} function. It
requires to specify a geodesic: \texttt{"PREM"} for a spherical Earth, or
\texttt{"WGS84"} for an elliptical one. A topography can be specified by
providing a path to global elevation data. Note that elevation data must be
given w.r.t. the sea level. Then, altitude values below zero -but above the
topography- are under the sea, if this option is activated. Note that the
topography path can encode a flat Earth model instead of a path to data, if it
starts as \texttt{"flat://"}. Then, the following \emph{float} value in the path
string is taken as the constant ground level. The constituent material of the
topography and its density can be customised as well. This might require to
modify the PUMAS Material Description File (MDF), though.

\subsection{Specifying a simulation context}
Each DANTON simulation requires to be executed within a specific
\dantondef{type}{struct\ danton\_context}. This allows to safely and
efficiently run multiple simulations in parallel, provided that \emph{lock} and
\emph{unlock} callbacks have been supplied. A new simulation context is
created (destroyed) with the \dantonfun{group/context}{danton\_context\_create}
(\dantonfun{group/context}{danton\_context\_destroy}) function. Then, the
simulation can be directly configured by modifying the structure field values.

The final $\tau$ state to sample is specified by providing a
\dantondef{type}{struct\ danton\_sampler} to the simulation context. This
sampler is created with the \dantonfun{group/sampler}{danton\_sampler\_create}
function. The desired final state properties are then specified by modifying the
exposed sampler data. Note that the final state must not necessarily be a
$\tau$. A neutrino flux can be sampled instead, or all $\tau$, $\nu$
particles inclusively. Once done, one must call the
\dantonfun{group/sampler}{danton\_sampler\_update} function in order to validate
the changes.

In order to run a Monte-Carlo simulation one must also provide one or more
primary $\nu$ flux model. This is done by attaching \dantondef{type}{struct\
danton\_primary} objects to the context, one per $\nu$ type. The API specifies a
generic primary object. The user might provide its own implementation, as long
as it conforms to the API, i.e. inherits from the base object. Two primary types
are provided by default. A discrete (constant) primary and a power-law spectrum.
They are created with the \dantonfun{group/discrete}{danton\_discrete\_create}
and \dantonfun{group/powerlaw}{danton\_powerlaw\_create} functions.

\subsection{Running a simulation}
A Monte-Carlo simulation is run with the \dantonfun{group/danton}{danton\_run}
function. This function takes a simulation context as argument. The number of
generated events must also be specified. In addition, one can request a specific
number of events to be sampled. Note that the number of sampled events is
generally lower than the number of generated events, even in a \emph{backward}
simulation. For example the backward sampled primary $\nu$ might have a too high
energy such that the event is rejected. In addition to running a forward or
backward Monte-Carlo, one can also operate a grammage scan of the geometry. This
is specified on the simulation context.

Simulated events conforming to the sampler specifications can be recorded. For
this purpose, a \dantondef{type}{struct\ danton\_recorder} must have been
attached to the context. The recorder is a generic API object as well. The user
might implement his own one. A \dantondef{group/text}{danton\_text} recorder is
provided by \texttt{libdanton}. It allows to dump the sampled events to a text
file. This recorder is created with the
\dantonfun{group/text}{danton\_text\_create} function. If an empty path is
provided for the output file, then the events are dumped to the standard output
(\texttt{stdout}).

\section{The DANTON executable}
A DANTON simulation can be run directly with the \texttt{danton} executable.
This executable encapsulates the \texttt{libdanton} API. Multithreading is
supported with \texttt{pthread}. The executable requires a steering file as
input, formated in JSON. Examples of steering files are provided with the
sources, in the \texttt{share/cards} folder. A steering file must contain a
single JSON object. The items of this object specify global configuration
parameters or categories of sub-parameters, as nested JSON objects. A list of
all parameters is given in table\,\ref{tab:cards}. In the following we provide
a brief summary of the available parameters. Note that most of these
directly reflect the \texttt{libdanton} API.

Currently (\texttt{v0.2.1}), the \texttt{danton} executable doesn't allow to use
another Physics than the default one specified in \texttt{libdanton}.
Nevertheless, the default Physics of \texttt{libdanton} can be changed at
compilation time by editing the folowing variables in the Makefile:
\texttt{DANTON\_DEFAULT\_PDF}, \texttt{DANTON\_DEFAULT\_MDF} and
\texttt{DANTON\_DEFAULT\_DEDX}.

\begin{table}[!t]
\begin{center}
\begin{tabular}{*3l}
\toprule
Key                       & Value                          & Description \\
\midrule
\texttt{"decay"}          & \texttt{boolean}               & If \texttt{true} the sampled taus are decayed. \\
\texttt{"events"}         & \texttt{integer}               & The number of Monte-Carlo events to run. \\
\texttt{"longitudinal"}   & \texttt{boolean}               & If \texttt{true} the transverse transport is disabled. \\
\texttt{"mode"}           & \texttt{string}                & The run mode, one of \texttt{"backward"}, \texttt{"forward"} or \texttt{"grammage"}. \\
\texttt{"output-file"}    & \texttt{string, null}          & The output file name or \texttt{null} for \texttt{stdout}. \\
\texttt{"requested"}      & \texttt{integer}               & The requested number of valid Monte-Carlo events. \\
\midrule
\multicolumn{3}{l}{\texttt{"earth-model"}} \\
\midrule
\texttt{\ \ "geodesic"}   & \texttt{string}                & The geodesic model: \texttt{"PREM"} (spherical) or \texttt{"WGS84"}. \\
\texttt{\ \ "sea"}        & \texttt{boolean}               & If \texttt{true} the PREM Earth is covered with sea. \\
\texttt{\ \ "topography"} & \texttt{[string, integer]}     & The topography data location and the in-memory stack size. \\
\midrule
\multicolumn{3}{l}{\texttt{"particle-sampler"}} \\
\midrule
\texttt{\ \ "altitude"}   & \texttt{float, float[2]}       & The altitude (range) of the sampled particles. \\
\texttt{\ \ "azimuth"}    & \texttt{float, float[2]}       & The azimuth angle (range) of the sampled particles. \\
\texttt{\ \ "elevation"}  & \texttt{float, float[2]}       & The elevation angle (range) of the sampled particles. \\
\texttt{\ \ "energy"}     & \texttt{float, float[2]}       & The energy (range) of the sampled particles. \\
\texttt{\ \ "latitude"}   & \texttt{float}                 & The geodetic latitude of the sampled particles. \\
\texttt{\ \ "longitude"}  & \texttt{float}                 & The geodetic longitude of the sampled particles. \\
\texttt{\ \ "weight"}     & \texttt{\{\$particle:float\}}  & The elevation angle (range) of the sampled particles.  \\
\midrule
\multicolumn{3}{l}{\texttt{"primary-flux"}} \\
\midrule
\texttt{\ \ "\$particle"} & \texttt{[\$model, \{...\}]}    & The primary spectrum model for the corresponding particle. \\
\midrule
\multicolumn{3}{l}{\texttt{\ \ \ \ "discrete"}} \\
\midrule
\texttt{\ \ \ \ "energy"}   & \texttt{float}               & The total energy of the primary. \\
\texttt{\ \ \ \ "weight"}   & \texttt{float}               & The weight of the primary, i.e. its integrated flux. \\
\midrule
\multicolumn{3}{l}{\texttt{\ \ \ \ "power-law"}} \\
\midrule
\texttt{\ \ \ \ "energy"}   & \texttt{float[2]}            & The energy range of the primary spectrum. \\
\texttt{\ \ \ \ "exponent"} & \texttt{float}               & The exponent of the power law. \\
\texttt{\ \ \ \ "weight"}   & \texttt{float}               & The weight of the primary, i.e. its integrated flux. \\
\midrule
\multicolumn{3}{l}{\texttt{"stepping"}} \\
\midrule
\texttt{\ \ "append"}     & \texttt{boolean}               & If \texttt{true}, append to the output file. \\
\texttt{\ \ "path"}       & \texttt{string}                & Path to the output file. \\
\texttt{\ \ "verbosity"}  & \texttt{integer}               & Verbosity level for recording Monte-Carlo steps. \\
\bottomrule
\end{tabular}
\end{center}
\caption{List of configuration parameters for \texttt{danton} steering cards.
\label{tab:cards}}
\end{table}

\subsection{Global simulation parameters}
Global simulation parameters are located at the top level of the steering file.
They allow to specify the operation mode: \texttt{"backward"},
\texttt{"forward"} Monte-Carlo or \texttt{"grammage"} scan. Scattering is
enabled by setting the \texttt{"longitudinal"} flag to \texttt{false}. The
default behaviour of the simulation is to print the sampled final states, on the
fly, to the standard output (\texttt{stdout}). The output stream can be
redirected to a file using the \texttt{"output-file"} global option.

\subsection{Earth model}
The Earth model parameters allow to configure the geometry of the simulation.
They are located under the \texttt{"earth-model"} root item. The
\texttt{"geodesic"} item allows to switch between a spherical Earth or an
elliptic one (WGS84). The legacy PREM has an external layer of 3\,km of sea
water. If the sea is disabled this layer is replaced with Standard Rock. Note
also that the \texttt{"topography"} item can encode a flat Earth model, as for
the library API. If global elevation data are provided instead, one must use the
WGS84 ellipsoid, i.e. a non spherical Earth.

\subsection{Particle sampler}
The particle sampler parameters allow to specify the final state(s) to sample.
They are located under the \texttt{"particle-sampler"} root item. Most of the
sampler parameters can be provided as a single float for a constant value or as
a list of two floats (\texttt{[min\_value, max\_value]}) for a range. The
\texttt{"weight"} JSON object specifies the particles to sample as a nested JSON
object. The valid particle names are \texttt{"nu\_tau"},
\texttt{"nu\_tau\textasciitilde"}, \texttt{"nu\_mu"},
\texttt{"nu\_mu\textasciitilde"}, \texttt{"nu\_e"},
\texttt{"nu\_e\textasciitilde"}, \texttt{"tau"} and
\texttt{"tau\textasciitilde"}.

\subsection{Primary $\nu$ flux}
The primary $\nu$ flux is provided as a JSON object, under the
\texttt{"primary-flux"} root item. For each item, the key must name a primary
particle while the value specifies the corresponding flux model. Valid names are
the same than previously, for the particle sampler. A flux model is given as a
list of two items. The first one is the type of the model, \texttt{"discrete"}
or \texttt{"power-law"}. The second one is a JSON object containing the model
parameters. Those are described in table\,\ref{tab:cards}.

\subsection{Recording Monte-Carlo steps}
The Monte-Carlo steps can be recorded, e.g. for visualisation purpose. This
is enabled by providing a \texttt{"stepping"} item at the root level of the
steering file. An output file name must be specified with the \texttt{"path"}
item. The verbosity level of recorded Monte-Carlo steps can be configured
with the \texttt{"verbosity"} flag. Setting it to zero results in all steps
being recorded. A higher value specifies a down-sampling period.

\section{Selected results}

\subsection{Validation of the Backward Monte-Carlo}
The BMC functionalities used by DANTON have been implemented as dedicated
packages, ALOUETTE\,\cite{ALOUETTE:GitHub}, ENT\,\cite{ENT:GitHub} and
PUMAS\,\cite{PUMAS:pages}. These packages have been tested independently. In the
present section, for purpose of illustration, we only present a few comparisons
of the end-to-end \emph{forward} and \emph{backward} chains, obtained with
DANTON. Specific cross-checks of PUMAS can be found in \cite{Niess:2018}.

\begin{figure}
  \centering
  \subfloat{\includegraphics[width=0.45\textwidth]{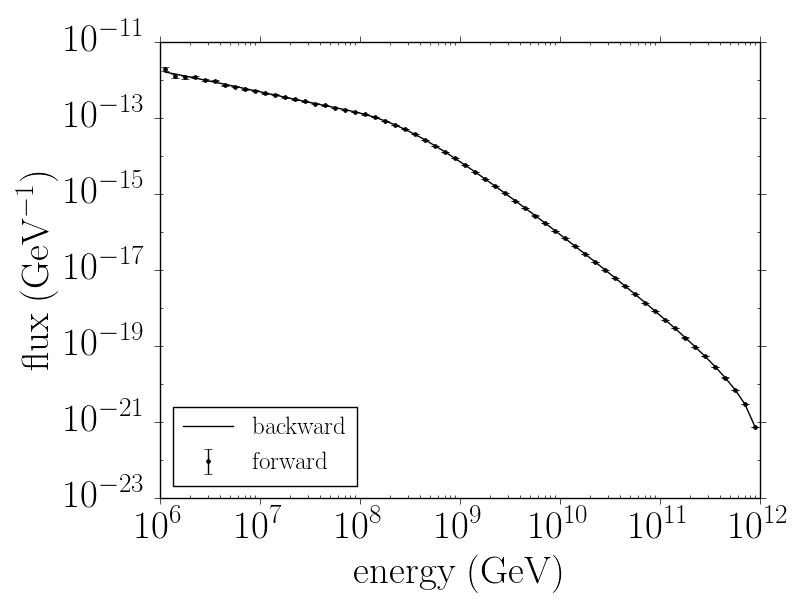}}
  \subfloat{\includegraphics[width=0.45\textwidth]{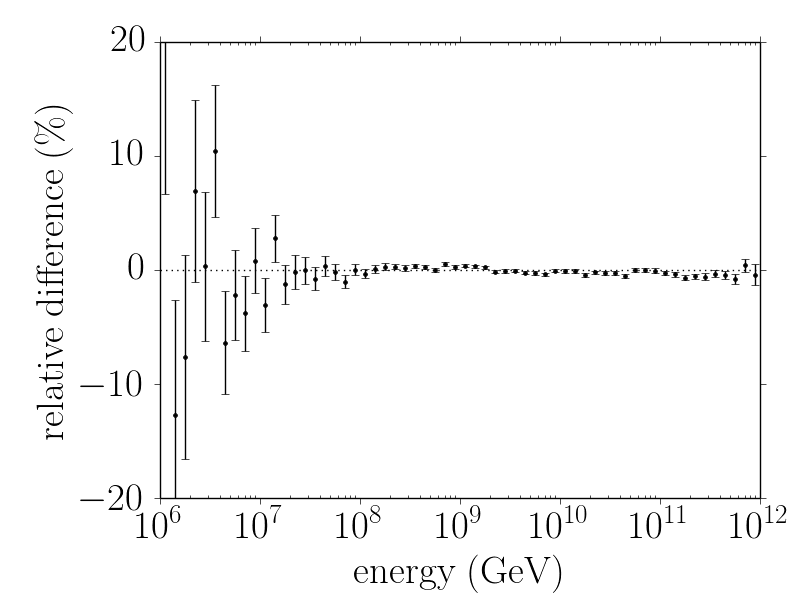}}
  \caption{Comparison of upward $\tau$ fluxes emerging from a PREM spherical
  Earth with an elevation angle of 1\,deg and for a $\frac{1}{E^2}$ primary
  $\nu_\tau$ flux. Left: absolute flux. Right: relative difference to the
  backward MC computation. \label{fig:flux_upward}}

  \vspace*{\floatsep}% https://tex.stackexchange.com/q/26521/5764

  \subfloat{\includegraphics[width=0.45\textwidth]{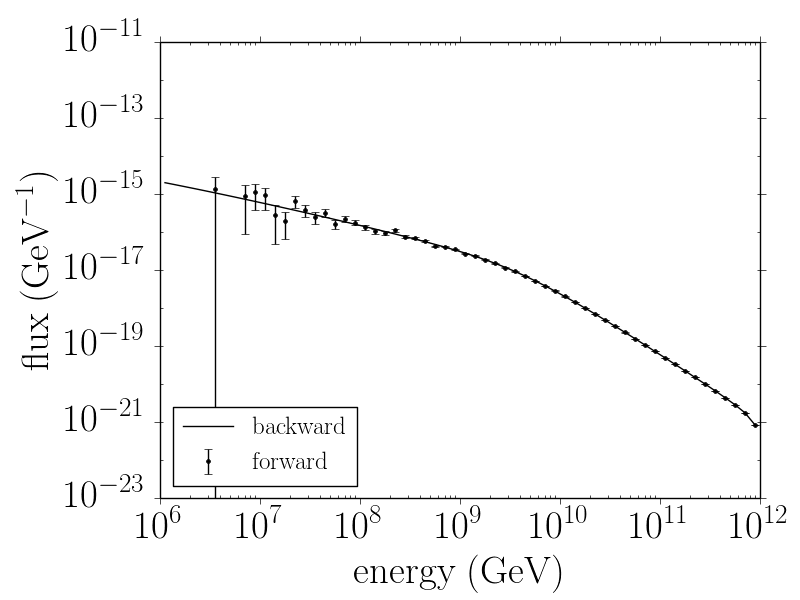}}
  \subfloat{\includegraphics[width=0.45\textwidth]{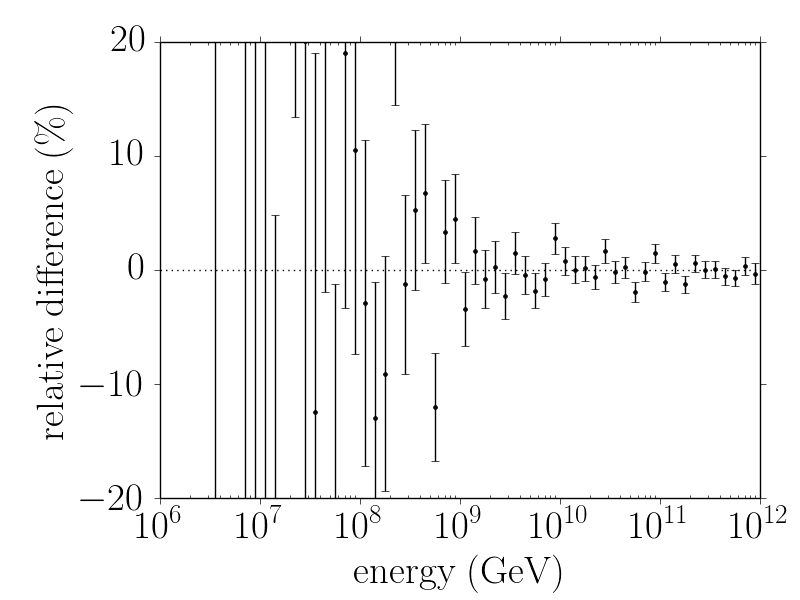}}
  \caption{Comparison of downward $\tau$ fluxes impacting the PREM spherical
  Earth with an elevation angle of -1\,deg and for a $\frac{1}{E^2}$ primary
  $\nu_\tau$ flux. Left: absolute flux. Right: relative difference to the
  backward MC computation. \label{fig:flux_downward}}

  \vspace*{\floatsep}% https://tex.stackexchange.com/q/26521/5764

  \subfloat{\includegraphics[width=0.45\textwidth]{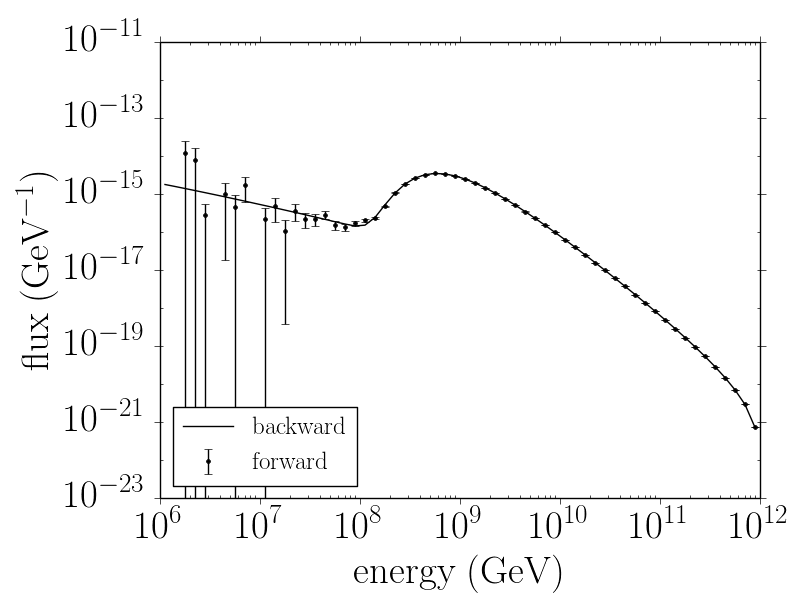}}
  \subfloat{\includegraphics[width=0.45\textwidth]{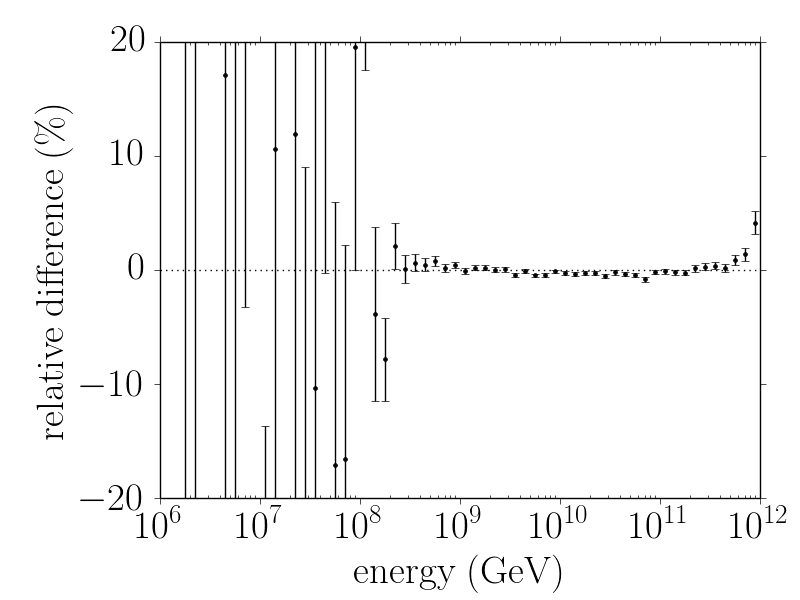}}
  \caption{Comparison of $\tau$ fluxes at 1 km above the PREM spherical Earth
  with an elevation angle of 1.4\,deg and for a $\frac{1}{E^2}$ primary
  $\nu_\tau$ flux. Left: absolute flux. Right: relative difference to the
  backward MC computation. \label{fig:flux_1km}}
\end{figure}

Figures~\ref{fig:flux_upward}, \ref{fig:flux_downward} and~\ref{fig:flux_1km}
show comparisons of the $\tau$ flux sampled above a spherical Earth, with a core
density described by the Preliminary Reference Earth Model (PREM, see Table~1 of
\cite{PREM}). It can be seen that the \emph{forward} and \emph{backward} results
agree within statistical errors. These results have been obtained with version
\texttt{v0.1.1} of DANTON and equivalent CPU time for both modes. The superior
efficiency of the \emph{backward} method is clearly visible here. Especially,
the \emph{backward} computed flux is much more accurate at both ends of the
enrgy range considered. Fig.~\ref{fig:flux_1km} is particularly interesting. In
this case the $\tau$ flux is sampled at 1\,km above the ground. Two components
are visible. Below $10^{17}\,\mathrm{eV}$ the $\tau$ flux is dominated by
grazing $\nu_\tau$ interacting in the atmosphere. One recovers the results of
fig.~\ref{fig:flux_downward} in this case. At higher energies, the $\tau$ flux
mostly results from $\nu_\tau$ interacting in the Earth crust, as in
fig.\ref{fig:flux_upward}. This flux is however suppressed at lower energies,
due to $\tau$ decaying in flight before reaching the altitude of 1\,km. Note
that in \emph{forward} mode, one hardly sees the first component, at lower
energies, whereas in \emph{backward} mode the whole spectrum is clearly
resolved.

\subsection{Comparison to NuTauSim}
NuTauSim and DANTON use similar ingredients but following different
implementations. NuTauSim was written in C++. It relies on parametrisations and
tabulations, e.g. for $\nu_\tau$ cross-sections, $\tau$ energy losses or $\tau$
decays. DANTON on the contrary glues together detailed (B)MC engines:
ALOUETTE\,\cite{ALOUETTE:GitHub}, ENT\,\cite{ENT:GitHub} and
PUMAS\,\cite{PUMAS:pages}. Below is a commented list of the main differences
between the two codes:

\begin{enumerate}[(i)]
        \item The $\tau$ energy loss is deterministic in NuTauSim. This is
        inaccurate at extreme energies, above $10^{20}\,\mathrm{eV}$. At these
        energies, fluctuations in the $\tau$ radiative energy losses become
        important, extending its range.

        \item NuTauSim allows to vary the cross-sections of physical processes
        within predefined models, using command line flags. DANTON, in principle
        allows any model to be pluged in, by changing the data set. However,
        this is less straightforward to do than in NuTauSim, since currently
        only one data set is packaged with DANTON.

        \item NuTauSim simulates a flat spherical Earth, as a modified version
        of the PREM\,\cite{PREM}. There is no atmosphere neither. DANTON allows
        for a detailed simulation of the (elliptical) Earth surface according
        to world wide Digital Elevation Models (DEM), e.g. ASTER\,\cite{ASTER}
        or SRTMGL1\,\cite{SRTMGL1}.

        \item NuTauSim supports only a single material: Standard Rock. DANTON
        has different materials, i.e. cross-sections, for air, rocks and water.
        Note that the $\nu_\tau$ cross-sections and $\tau$ energy loss in rocks
        and water can differ by 10\,\%.

        \item NuTauSim is a purely longitudinal MC. The scattering angle of the
        $\tau$ w.r.t. the primary $\nu_\tau$ direction is not simulated.

        \item NuTauSim is 1\,kLOC, versus 15\,kLOC for DANTON, including its
        dependencies. DANTON is slower and harder (more error prone) to develop.
\end{enumerate}

Detailed comparisons have been performed between the two codes. For these
comparisons, NuTauSim was configured with \emph{standard} $\nu_\tau$
cross-sections (\texttt{CCmode=0}, i.e. the middle parametrisation of
\citet{Connolly:2011}) and $\tau$ energy loss (\texttt{ALLM(0)}). NuTauSim was
hacked as well in order to use the standard PREM, as initially defined
by~\cite{PREM} and used in DANTON. Indeed, NuTauSim implements a modified
version of the PREM with a 7\,km larger Earth radius. Note also that version
\texttt{v0.1.1} of DANTON was used for the comparison, i.e. not the latest one.
In particular, the CT14\_NNLO PDF have been used, instead of the CT14\_NLO PDF
in the current \texttt{v0.2.1} version.

\begin{figure}[!t]
  \centering
  \subfloat{\includegraphics[width=0.48\textwidth]{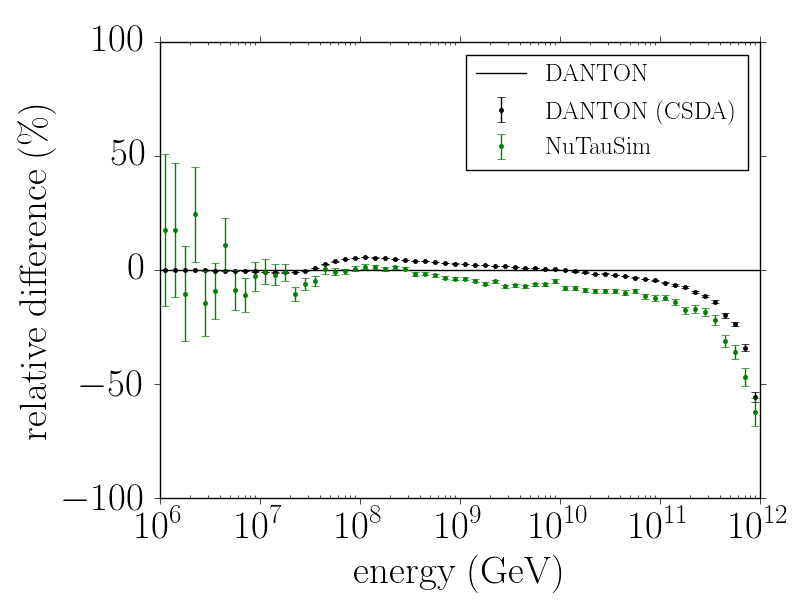}}
  \subfloat{\includegraphics[width=0.48\textwidth]{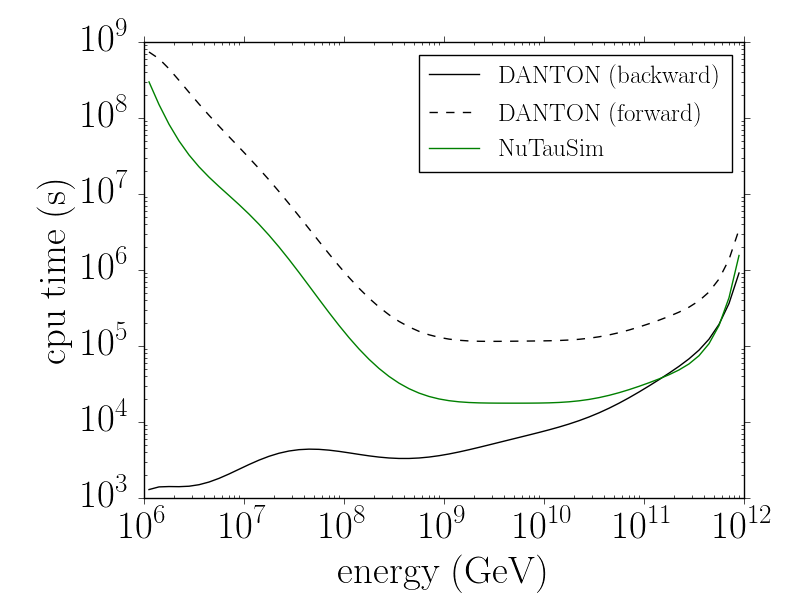}}
  \caption{Comparison of NuTauSim and DANTON for the upward $\tau$ flux
  emerging from the Earth with an elevation angle of 1\,deg and for a
  $\frac{1}{E^2}$ primary $\nu_\tau$ flux. Left: relative difference to DANTON.
  Right: CPU time needed for reaching a 1\,\% Monte-Carlo accuracy.
  \label{fig:flux_nts}}
\end{figure}

As an illustration, the comparison of the the $\tau$ flux emerging from the PREM
spherical Earth, with 1\,deg of elevation, is shown on fig.\ref{fig:flux_nts}.
Significant differences can be observed above $10^{20}$\,eV. These differences
are due to fluctuations in the $\tau$ energy loss, as discussed previously. This
was cross-checked by configuring PUMAS to use the Continuous Slowing Down
Approximation (CSDA) instead of the detailed simulation for $\tau$ energy
losses. Then, DANTON and NuTauSim yield very similar results, with a 10\,\%
systematic offset. This offset was checked to be consistent with the differences
in the $\nu_\tau$ cross-sections and $\tau$ energy loss models used by the two
codes. Other angles of emergence have been investigated as well. Overall, when
using the same geometry, DANTON and NuTauSim are in very good agreement, within
10\,\%. However, when using the default NuTauSim geometry, the emerging $\tau$
fluxes can differ by a factor of 10. The higher the $\tau$ energy, the higher
the discrepancy.

The right part of fig.\ref{fig:flux_nts} also shows the CPU time needed in order
to reach a 1\,\% Monte-Carlo accuracy on the emerging flux. Note that in BMC,
the Monte-Carlo events are necessarily weighted. Consequently, it is not
relevant to compare the CPU time per Monte-Carlo event, since events are not
equally important. Therefore, we instead compare the CPU time needed in order to
reach a given accuracy, 1\,\% in this case. It can be seen that in
\emph{forward} Monte-Carlo mode DANTON is significantly slower than NuTauSim, up
to a factor of 10. This is not surprising considering the differences in
complexity and level of details of the two codes. For example, DANTON implements
a detailed stepping through the geometry, while NuTauSim maps the density
distribution at initialisation. Indeed, in NuTauSim all Monte-Carlo events have
the same initial direction and they don't scatter. Therefore, they all go
through the same density distribution.

In BMC mode, DANTON is more effcient than NuTauSim at lower energies, and
equally efficient at extreme energies. Hence, the overburden resulting from
doing a detailed simulation is counterbalanced by the gain in efficiency
resulting from the backward sampling.

\section{Conclusion}
In this brief technical note we presented a preliminary version
(\texttt{v0.2.1}) of the DANTON Monte-Carlo package. The core functionalities of
DANTON have been implemented as a dedicated library: \texttt{libdanton}. This
allows a direct integration of DANTON in a higher level simulation. The DANTON
package also provides an executable: \texttt{danton}, encapsulating
\texttt{libdanton}. With this executable one can run DANTON simulations from the
command line, according to steering files in JSON format.

A dedicated \emph{backward} Monte-Carlo procedure allows DANTON to achieve high
efficiency while performing a detailed simulation of the Physics and of the
Earth topography. The validity of the procedure was cross-checked by comparison
to classical Monte-Carlo results. When using identical geometries, good
agreement was found with the NuTauSim simulation code, within theoretical
systematics on the $\nu_\tau$ and $\tau$ cross-section models.

\bibliographystyle{apsrev}
\bibliography{main}

\end{document}